\documentclass[fleqn,10pt]{SelfArx} 

\definecolor{color1}{RGB}{0,0,90} 
\definecolor{color2}{RGB}{0,20,20} 

\usepackage{hyperref} 
\hypersetup{hidelinks,colorlinks,breaklinks=true,urlcolor=color2,citecolor=color1,linkcolor=color1,bookmarksopen=false,pdftitle={Title},pdfauthor={Author}}

\Archive{Open Source Toolkit} 

\PaperTitle{MiniConf - A Virtual Conference Framework} 

\Authors{Alexander M. Rush\textsuperscript{1}, Hendrik Strobelt\textsuperscript{2}} 
\affiliation{\textsuperscript{1}\textit{CS+Cornell Tech, Cornell University, New York NY, USA}} 
\affiliation{\textsuperscript{2}\textit{MIT-IBM Watson AI Lab, IBM Research, Cambridge MA, USA}} 
\affiliation{\textbf{Correspondence}: info@mini-conf.org} 

\newcommand{\keywordname}{Keywords} 

\Abstract{MiniConf is a framework for hosting virtual academic conferences motivated by the sudden inability for these events to be hosted globally. 
The framework is designed to be global and asynchronous, interactive, and to promote browsing and discovery. We developed the system to be sustainable and maintainable, in particular ensuring that it is open-source, easy to setup, and scalable on minimal hardware. In this technical report, we discuss design decisions, provide technical detail, and show examples of a case study deployment. 
}

\begin{document}

\flushbottom 

\maketitle 
\thispagestyle{empty} 


\section*{Introduction} 

\addcontentsline{toc}{section}{Introduction} 

Conferences are a critical aspect of the academic research communities, but they also require significant organizational resources and participant buy-in to host effectively. In recent years there has been a call for alternatives to physical conferences motivated by the impact of travel on climate change, difficulty for parents and caregivers, and increasing challenges of obtaining travel visas. This demand became suddenly pressing with the worldwide onset of COVID-19, and the shocking collapse of the ability to hold physical conferences at all~\cite{Woolston2020-qx}. This catalyst has forced many communities to construct new methods and approaches for holding academic events. 

Motivated by these challenges, we were pressed to develop a minimal virtual conference system. While it is impossible to fully reconstruct an academic conferences in a virtual space, we aimed for capturing three key elements:  scheduling across borders,  facilitating interactivity,  promoting browsing and exploration. These demands were particularly relevant to the computer science community and its paper style; however, many of these properties are shared by other conference types.

In addition to these funcitonal demands, a key concern was developing a sustainable system for academic communities. While there are many advanced proprietary applications for running virtual events, it was important for us that the system be fully open-source, easily deployable by someone with little-to-no web development experience, and scalable to virtually any size audience on low-cost hardware. 

Our system, MiniConf, is an attempt to satisfy this set of goals. The main system provides a way to easily configure a conference with a global schedule, plug-and-play interativity tools, and out of the box visual browsing and search. The backend is completely setup using only flat-file settings, usable with almost no programming, and deploys as a completely static site that can scale to the largest academic events. The code and instructions are available at:  \url{http://mini-conf.org}. As of this technical report the system has been successfully employed at several major academic conferences include ICLR, ACL, and AKBC, as well as many smaller events.


\paragraph{Related Work}

There are many proprietary hosted tools for hosting an online virtual conference. These include conference management tools such as Whova, fully virtualized environments like 6connex, and video hosting platforms such as SlidesLive. Within the computer science academic community, there is a tradition of utilizing open-source communication tools as an alternative to proprietary software. This allows the community to maintain control of the 
conference process. For instance, in the area of review management tools like OpenReview and HotCRP allow for conferences to control aspects of the review process. MiniConf aims to bring this to conference hosting. While there are many other open-source tools that are used for website development, MiniConf aims to provide easy defaults and deployment targeting only virtual conferences. 


\section{Case Study}

MiniConf was originally designed to host the International Conference of Learning Representations (ICLR 2020) as a virtual conference. While every conference is different, ICLR is useful as a case study for the challenges that required a new tool. We highlight three general challenges behind the conference. 

\subsection{Scheduling and International Participation}
The conference was planned to have participants from all over the world who would be participating within their local time zone. The virtual conference would need to provide clear scheduling and asynchronous events. Furthermore the website would need to be accessible from a large number of devices and work in many difference countries.

\subsection{Dynamic Interaction}

At the same time, a main aspect of a conference is the social interactions that occur both in official and unofficial channels. For the conference to be successful there would need to be Q\&A sessions, social sub-group gathering, or spontaneous. These aspects would need to be integrated within the conference website while also allowing for dynamic changes.  

\subsection{Browsing and Discovery}

A key aspect of the conference was the presentation of work through ``posters''. Each paper presentation was instead presented through a recorded video, but these needed to be presented in a clear and organized way, while also allowing for casual browsing as well as directed search.

\section{System Design}

The MiniConf software is a minimal system to host a workshop or conference. It provides components for event schedules, exploration of papers and posters, list of social events, and additional options to communicate through chat or FAQs. 

Before discussing the tools provided by the system, we give an overview of the main technical design decisions: 

\begin{itemize}[]
    \item It has a \textit{frozen} architecture that creates a \textbf{static site}. This decision allows the conference to be deployed on basically any web server while scaling easily and cheaply. The backend is run only by the developer before deployment. During the conference, nothing is exposed that could become a security vulnerability. 
    \item It can be completely configured without a database using only \textbf{flat files} such as CSV or YAML. While this prevents the ability to scale to massive events, it greatly lowers the threshold for setting up the system and reduces the number of technologies for the conference organizer.  
    \item The front-end is \textbf{easy to extend} through HTML to allow for  additional components. While a standard setup can be achieved in short time, we also want to enable customization for conferences with more time budget and human resources.  
\end{itemize}

Based on these underlying design principles, the framework includes a skeleton with the following components.

\subsection{Introductory Materials}

\begin{figure}[tb]
    \centering
    \includegraphics[width=\linewidth]{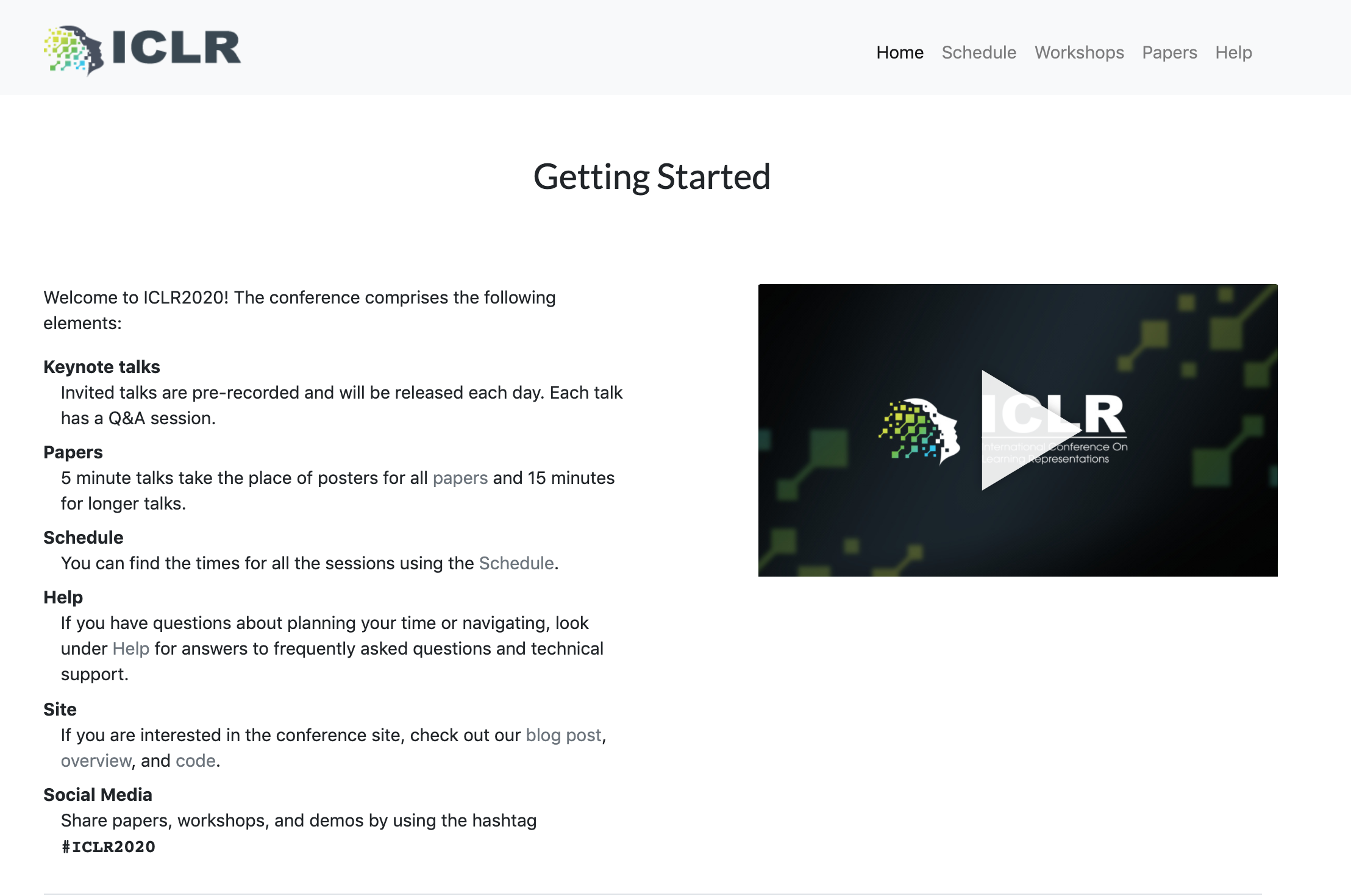}
    \caption{Landing page example.}
    \label{fig:landing_page}
\end{figure}

As virtual conferences are still a new event for most participants, the landing page makes it easy to provide basic information about the conference and act as a quick start guide for attendees. It should inform them about different event types of the conference and how they are reflected in the webpage. An video intro allows for a more personal welcome and allows to explain the intended character of the gathering.
MiniConf leaves the design for the welcome part flexible, but supports easy extension such as organizers, sponsors, welcome videos, or other material.


\subsection{Event Schedules}
MiniConf has a central schedule component that links out to: (1) event pages for social getherings or keynotes and (2) to browsing subsets of papers for presentation sessions. The schedule utilizes the attendee's timezone by default but allows adjustment. A daily view of events has been shown to be useful. While running it during ICLR, the calendar view has been the most visited component of MiniConf. Therefore, we recommend to keep it as structured as possible and consider it as the major navigation element. 
The calendar also can be provided as flat file such as an ICal file. For conference organizers, this schedule can be created utilizing shared calendar tools for graphical scheduling.


\begin{figure}[tb]
    \centering
    \includegraphics[width=\linewidth]{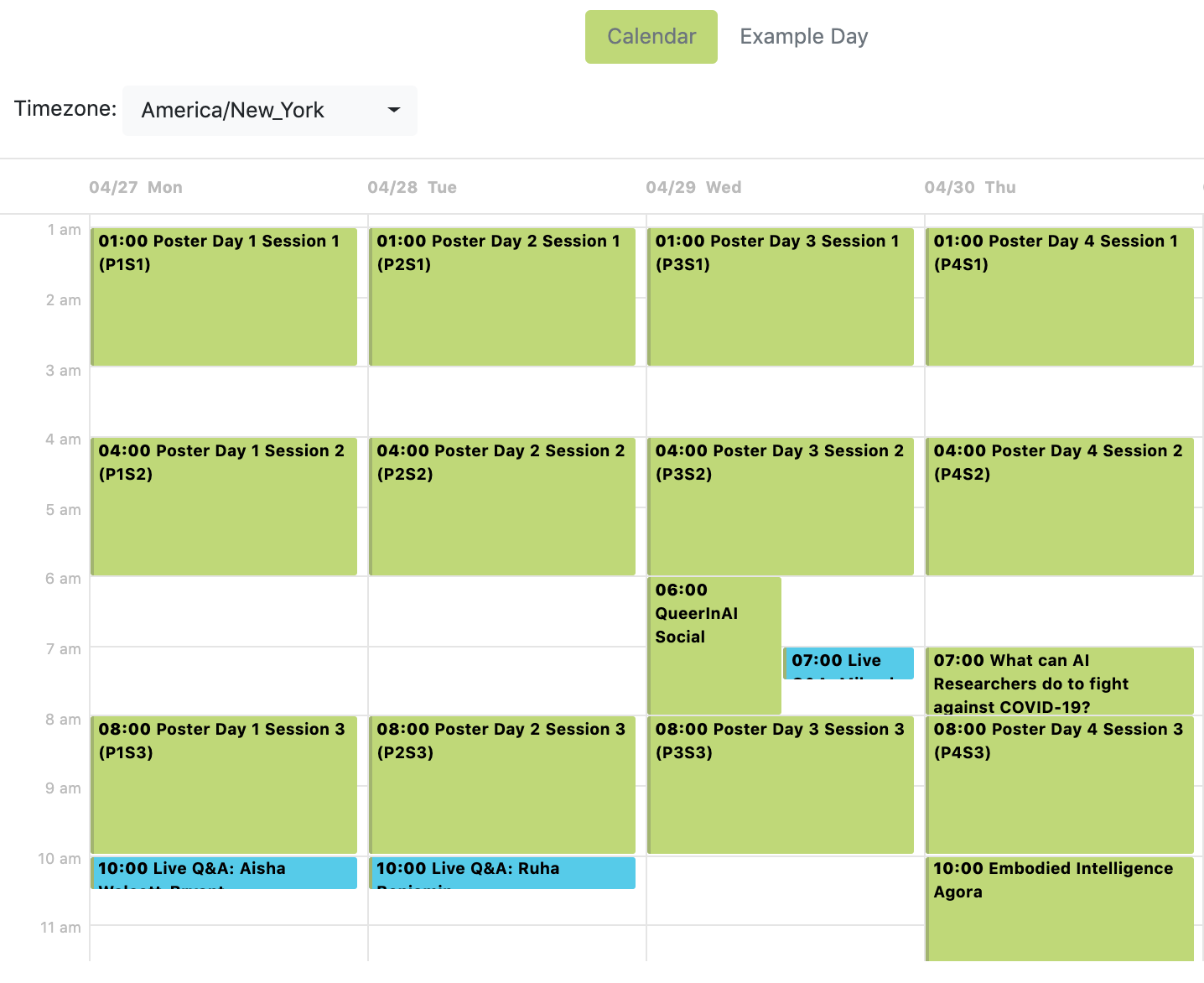}
    \caption{Calendar component.}
    \label{fig:cal}
\end{figure}

\subsection{Browsing Papers}
\begin{figure}[tb]
    \centering
    \includegraphics[width=\linewidth]{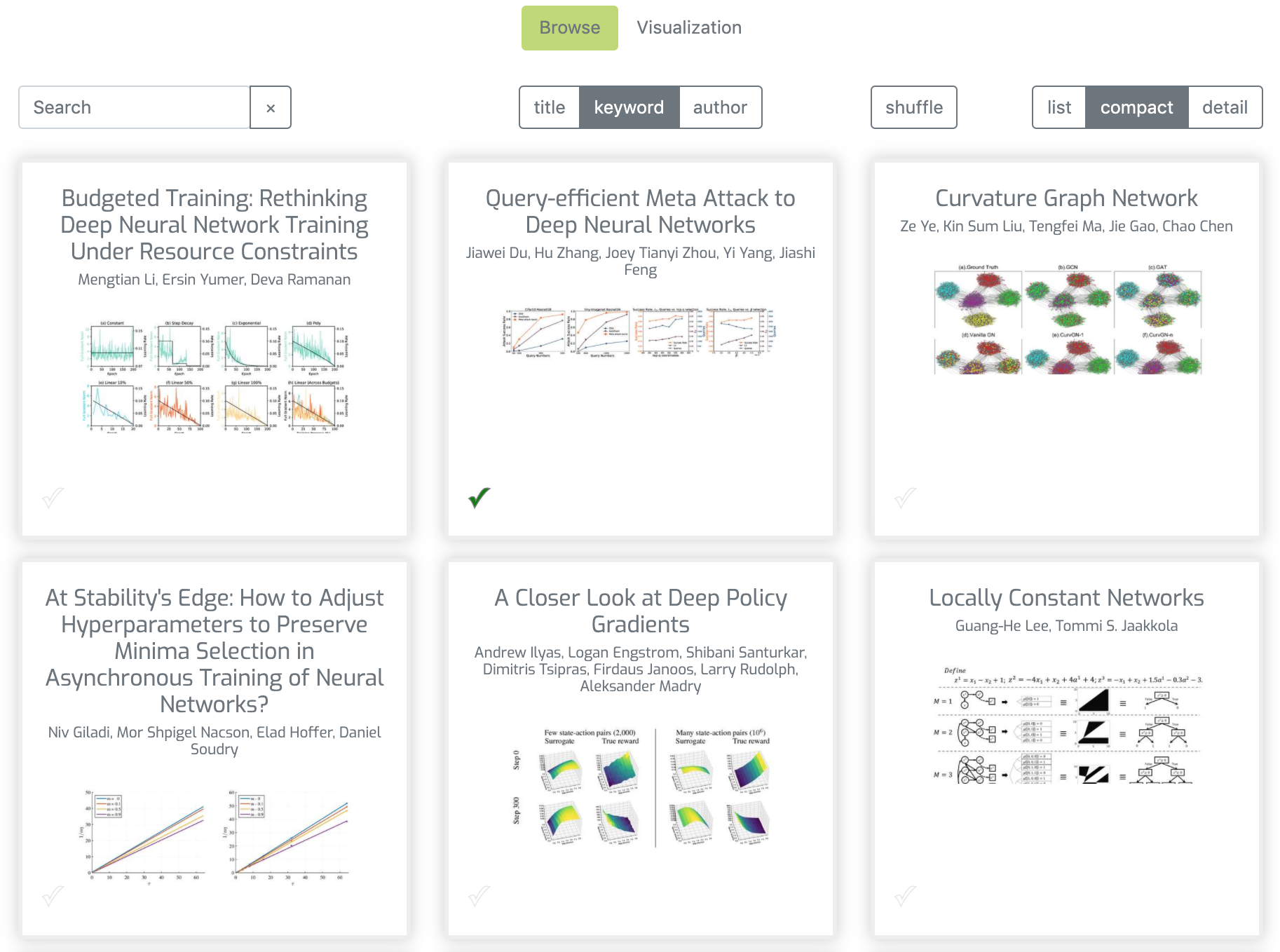}
    \caption{Paper browser.}
    \label{fig:browsing}
\end{figure}

Paper browsing should support search features that allow for targeted search but also support serendipitous browsing. The interface in \autoref{fig:browsing} allows faceting all available papers by authors, title, and keyword to allow targeted search. To facilitate exploration it randomly shuffles papers and user can trigger a re-shuffle. Three levels of details provide the user to adjust the trade-off between number of papers shown on the screen vs. granularity of information. The \textit{list} level only shows title and author, the \textit{compact} level adds an interesting figure, and \textit{details} provides abstract and keywords. Each visited paper is automatically marked by checkmark but can be unmarked if the user intends so. The interface is constructed on the frontend, based on a cached CSV file of the conference. Again the goal is to allow the conference organizers to directly change things using only standard office tools.


\begin{figure}[tb]
    \centering
    \includegraphics[width=\linewidth]{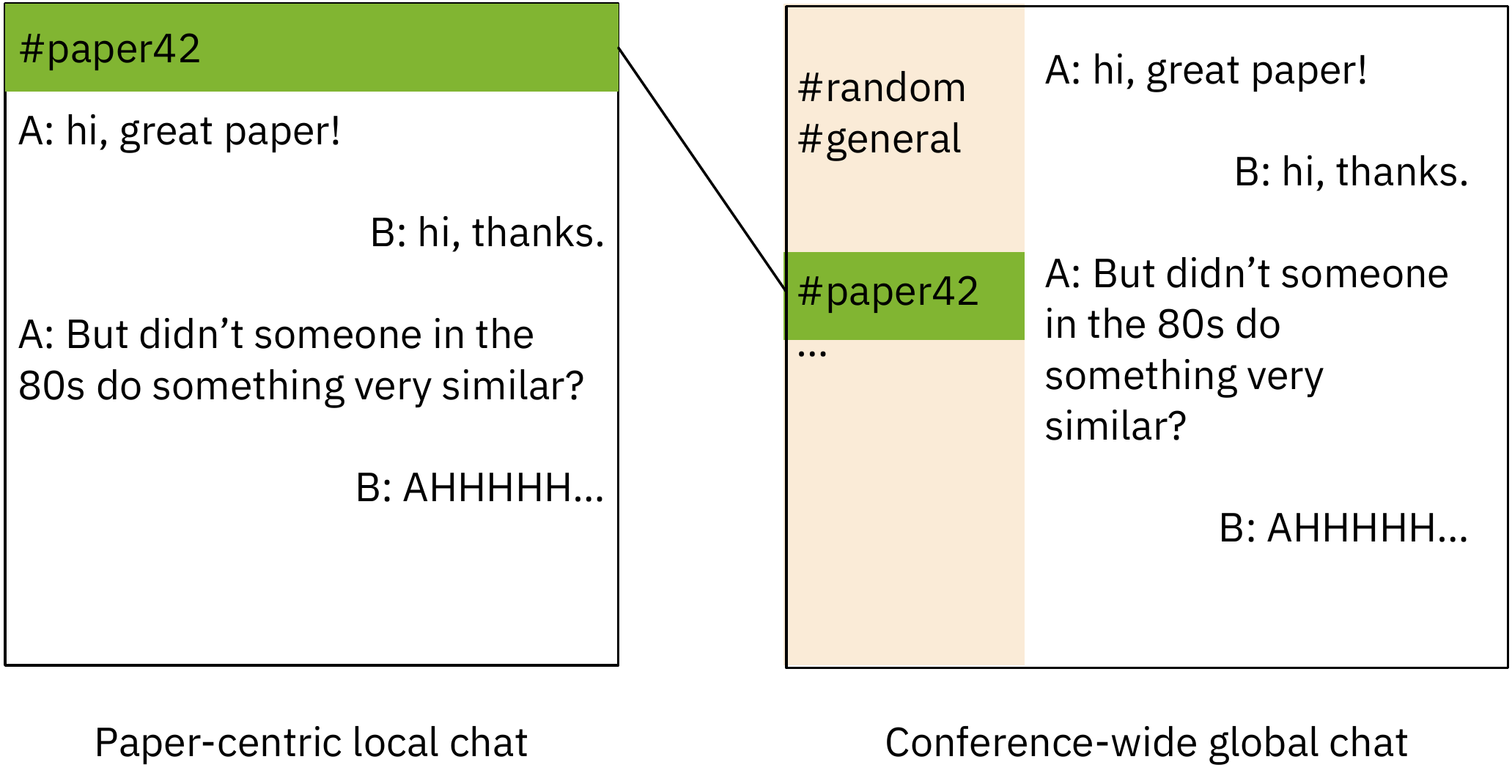}
    \caption{Interactivity for global chat and local paper comments.}
    \label{fig:chat}
\end{figure}

\subsection{Visual Exploration of Papers}

\begin{figure*}[tb]
    \centering
    \includegraphics[width=\linewidth]{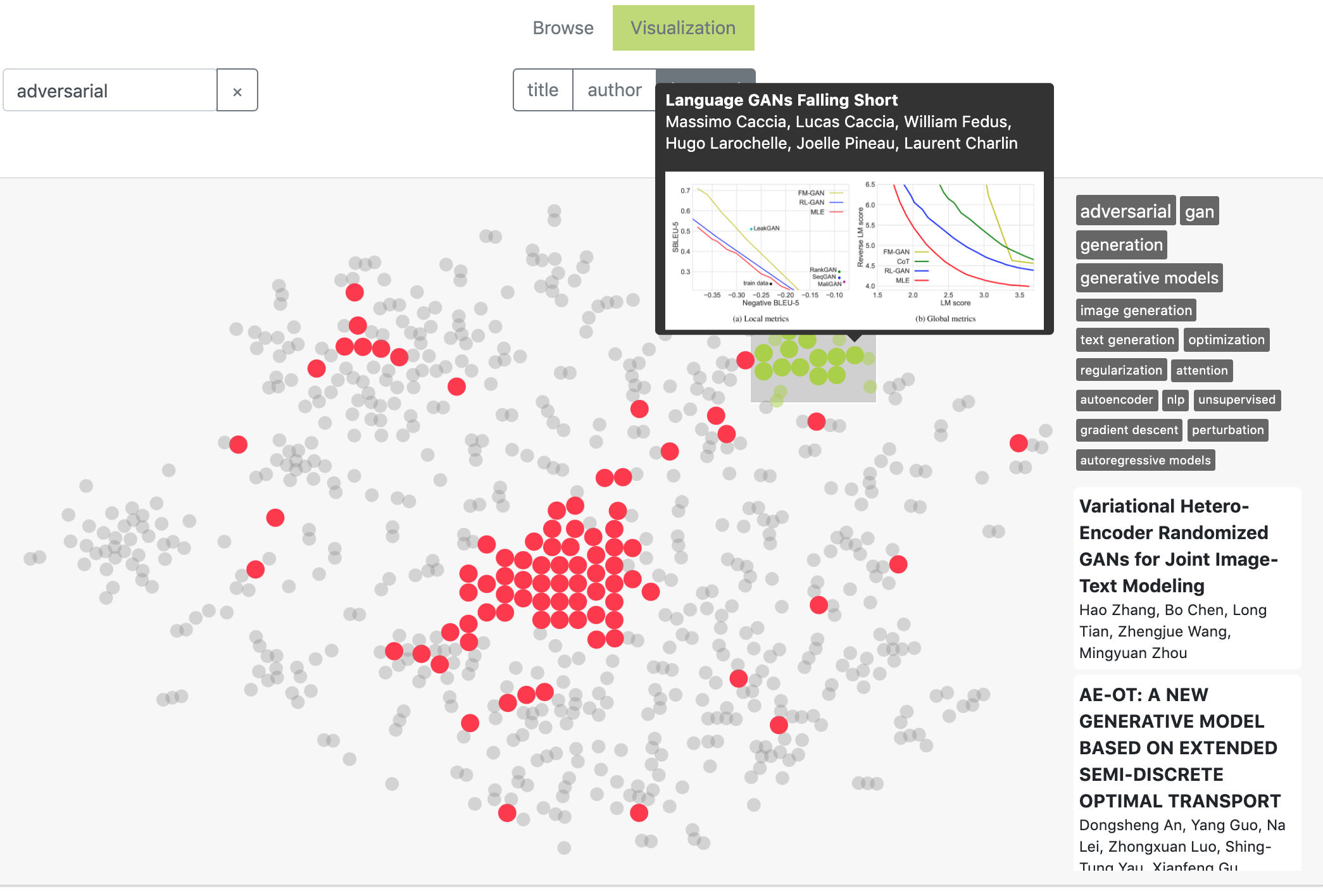}
    \caption{Visualization of papers. Each dot represents a document and hovering over a dot reveals key information the paper (title, author, image). Dots are placed by applying a projection technique (tSNE) on document embeddings.  Documents can be faceted and appear in red if the match a filter criterion. The user can select a select a subset of documents and gets presented an aggregation of keywords describing the selection.}
    \label{fig:paper_vis}
\end{figure*}

To provide a comprehensive overview of the diversity of papers, MiniConf provides an interactive visualization component that represents papers as meaningfully arranged dots on a canvas. Each dots position is derived from a projection on high-dimensional document embeddings for all papers. Hovering over a dot reveals the key attributes equal to the \textit{detail} level described in document browsing.

To highlight papers in red, the same facets for author, name, keyword can be applied. 
To investigate groups or cliques of dots, the user can create selection by drawing a selection box. The combination of facet highlighting and user selection allows powerful subgroup analysis. In \autoref{fig:paper_vis}  the user searched for the keyword \textit{adversarial } and two clusters pop up in red. By creating a selection box on top of each two clusters and comparing the aggregated keywords, the user can  conclude that one cluster refers to papers about \textit{Generative Adversarial Networks (GANs)} whilst the other cluster refers to \textit{adversarial robustness/attacks}. Both clusters relate to the concept of \textit{adversarial} but disambiguate it different contexts. 

To prepare for the visualization component, MiniConf provides one general but interchangeable pipeline to derive document embeddings and infer positions for each document (details in \autoref{sec:Methods}). 


\subsection{Paper Presentation and Interactivity}
Each paper has a dedicated own page to display all related data in the most detailed form including but not limited to: videos of presentations, links to live sessions, and rendering of papers or posters. Additionally, each paper is assigned a chat room in a global conference-wide chat system. While the main page is static, that interactive chat room is embedded directly on the page to allow for real time comments and discussion. Each of these chat rooms can also be accessed through a global communication system. Other dynamically created chat rooms can be created by users through the conference. The embedding of this single chat group on each paper detail page helps to integrate a robust solution for local and global communication using the same technology.

\section{Automatic Tools}
\label{sec:Methods}

The main MiniConf framework uses simple inputs to create a customizable conference site. It can be adapted to any conference backend or setup manually. In the process of developing the system, we found that there were some general purpose aspects that were common to many conferences. We therefore included a set of automatic tools, included as scripts,  developed to help better present information in the conference format.

\subsection{Image Extraction}
The main paper browsing mode of the conference presents each paper to the attendee in the form of a document card with title and authors. However, we found that in a sea of papers with posters it is difficult to distinguish these cards from each other (see also \cite{Strobelt09}). We therefore set up a system for extracting images from submitted papers. We utilized a deep learning based system for image extraction utilizing the PubLayNet dataset and pretrained model \cite{zhong2019publaynet}.
For ICLR, we found this method to be effective enough that we could use it out of the box with little problems.


\subsection{Document Embedding}

Several applications in MiniConf require the use of vector embeddings representing the papers displayed. To compute these embeddings, MiniConf 
uses the abstracts of articles and a pretrained sentence embedding model~\cite{Wieting2019SimpleAE}. This procedure was based on a method 
for assigning reviewers to papers. In MiniConf, these embeddings can be used as a basis for visualization, browsing, and papers recommendation.

\subsection{Chat/Auth Integration}

MiniConf is a static website, but it includes scripts for dynamic integration with authentication platforms and with global chat systems. Most deployments utilize an open-source chat server. MiniConf includes backend scripts for populating local paper channels as well as front-end integration points for both local channels and global integration into the main site. In addition, both the chat server and the site itself can be gated using external authentication servers. 

\section{Conclusion}

Conferences are critical to academic communities. It seems both unlikely and undesirable that virtual events will fully replace live events. However, virtual conferences can help supplement and, in extreme times, act as stand-ins. MiniConf is open-source software designed to make it easy and affordable to host one of these events. It has now been tested in several major conferences, and succeeded in providing async scheduling, dynamic interactions, and useful browsing and search. It continues to be extended with new features and with sample conferences as virtual events are hosted around the globe. 




\phantomsection
\section*{Acknowledgments} 
\addcontentsline{toc}{section}{Acknowledgments} 

Thanks to Darren Nelson for the original design sketches. Shakir Mohamed, Martha White, Kyunghyun Cho, Yoshua Bengio, Hugo Larochelle, Lee Campbell, and Adam White for planning and feedback. Hao Fang, Junaid Rahim, Jake Tae, Yasser Souri, Soumya Chatterjee, and Ankshita Gupta for contributions.

\clearpage
\phantomsection
\bibliographystyle{unsrt}
\bibliography{miniconf}


\end{document}